\documentclass[12pt]{article}

\usepackage{natbib}
\usepackage{amsmath}
\usepackage[dvipsnames]{xcolor}
\usepackage{graphicx,psfrag,epsf}
\usepackage{enumerate}
\usepackage{algorithm}
\usepackage{algorithmicx}
\usepackage{algpseudocode}
\usepackage{url} 
\usepackage{mathtools,amssymb,amsthm}
\usepackage{soul} 
\usepackage{float} 
\usepackage[colorlinks=true,linkcolor=blue,urlcolor=NavyBlue,citecolor=RubineRed]{hyperref}
\usepackage{booktabs}
\usepackage{multirow}
\usepackage{kotex}
\usepackage{color, colortbl}
\usepackage{caption}
\usepackage{floatpag}

\usepackage{todonotes} 
\usepackage{comment} 

\newif\ifshow 
\showfalse 

\ifshow
  \includecomment{wrap}
\else
  \excludecomment{wrap} 
\fi

\newcommand{\beq}{\begin{eqnarray}}
\newcommand{\eeq}{\end{eqnarray}}

\newcommand{\blambda}{\boldsymbol{\lambda}}

\def\spacingset#1{\renewcommand{\baselinestretch}%
{#1}\small\normalsize} \spacingset{1}

\theoremstyle{plain}

\theoremstyle{definition}

\newcommand{\titlefont}{\fontsize{17}{22}\selectfont\bfseries}

\begin{document}

\title{\titlefont High-Dimensional Mediation Analysis for Generalized Linear Models Using Bayesian Variable Selection Guided by Mediator Correlation}

\author{Youngho Bae$^{1 \dag}$, Chanmin Kim$^{1 \dag}$, Fenglei Wang$^{2}$, Qi Sun$^{2,3,4}$, Kyu Ha Lee$^{2,3,5\ast}$\\ \\
    \textit{\small $^{1}$Department of Statistics, Sungkyunkwan University, Seoul, South Korea}\\
    \textit{\small $^{2}$Department of Nutrition, Harvard T.H. Chan School of Public Health, Boston, MA, U.S.A.}\\
    \textit{\small $^{3}$Department of Epidemiology, Harvard T.H. Chan School of Public Health, Boston, MA, U.S.A.}\\
    \textit{\small $^{4}$Channing Division of Network Medicine, Brigham and Women's Hospital, Boston, MA, U.S.A.}\\
    \textit{\small $^{5}$Department of Biostatistics, Harvard T.H. Chan School of Public Health, Boston, MA, U.S.A.}\\
    {\small $^\dag$co-first authors}\\   
    {\small $^\ast$klee@hsph.harvard.edu}}
    
    \date{}
    
\maketitle

\begin{abstract}
\noindent High-dimensional mediation analysis aims to identify mediating pathways and to estimate indirect effects linking an exposure to an outcome. In this paper, we propose a Bayesian framework to address key challenges in these analyses, including high dimensionality, complex dependence among omics mediators, and non-continuous outcomes. Furthermore, commonly used approaches assume independent mediators or ignore correlations in the selection stage, which can reduce power when mediators are highly correlated. Addressing these challenges leads to a non-Gaussian likelihood and specialized selection priors, which in turn require efficient and adaptive posterior computation. Our proposed framework selects active pathways under generalized linear models while accounting for mediator dependence. Specifically, the mediators are modeled using a multivariate distribution, exposure-mediator selection is guided by a Markov random field prior on inclusion indicators, and mediator-outcome activation is restricted to mediators supported in the exposure-mediator model through a sequential subsetting Bernoulli prior. Simulation studies show improved operating characteristics in correlated-mediator settings, with appropriate error control under the global null and stable performance under model misspecification. We illustrate the method using real-world metabolomics data to study metabolites that mediate the association between adherence to the Alternate Mediterranean Diet score and two cardiometabolic outcomes.

\end{abstract}



\spacingset{1.45} 



\section{Introduction}

Mediation analysis addresses a different question than the total causal effect alone: it asks how the effect of an exposure or treatment on an outcome is transmitted through intermediate variables. In this framework, a variable known as the mediator is assumed to lie on the causal pathway between exposure and outcome. The total causal effect can then be decomposed into an indirect effect operating through the mediator, and a direct effect, the portion that bypasses the mediator and influences the outcome directly, mediation analysis provides deeper insight into causal mechanisms. This decomposition is especially informative when the direct and indirect effects act in opposite directions, since the total effect may be small and can appear null even when meaningful pathways exist. In such cases, mediation analysis becomes essential for revealing the underlying, counteracting pathways that standard causal inference methods fail to detect.

Within the potential outcomes framework, a wide range of methodologies have been developed for mediation analysis under a single mediator scenario. These methods typically rely on the sequential ignorability assumption (a two-stage extension of the standard strong ignorability assumption) \citep{imai2010identification}. Classical approaches include linear structural equation modeling \citep{baron1986moderator, mackinnon1994analysis}, while more recent advancements range from nonparametric identification \citep{imai2010identification} and semiparametric modeling \citep{tchetgen2012semiparametric} to Bayesian approaches \citep{kim2017framework,kim2018bayesian}.

When multiple mediators are present, however, the identification of causal mediation effects requires new identifying assumptions and structural considerations. Mediators may operate through independent pathways, or they may exhibit sequential dependence in which the effect flows through mediators in a prespecified order. \cite{imai2013identification} proposed identification assumptions that accommodate both causally independent and sequentially dependent multiple-mediator settings, leading to a variety of estimation strategies \citep{vanderweele2014mediation,daniel2015causal,steen2017flexible}. Building on this foundation, \cite{kim2019bayesian} developed a Bayesian framework that allows for dependence among mediators while flexibly estimating their marginal distributions. 

In many contemporary scientific applications, researchers face settings in which the number of potential mediators is large, as in genomic, metabolomic, and microbiome studies. In such high-dimensional mediation problems, the primary inferential goals are not only to accurately quantify indirect effects but also to identify the relatively small subset of mediators that constitute the true causal pathways. Variable-selection strategies based on penalized regression \citep{zhang2016estimating,gao2019testing,perera2022hima2,zhang2022high} have been widely employed for this purpose. However, penalty-based procedures typically treat selection and estimation as separate stages, which complicates uncertainty quantification after selection. Alternatively, joint modeling approaches \citep{aung2020application,song2020bayesianOmics,song2021bayesian} simultaneously estimate the exposure–mediator and mediator–outcome relationships, thereby providing coherent uncertainty quantification for indirect effects. Although these methods address selection-induced uncertainty, they often assume independence among mediators, an assumption that is rarely justified in biological or environmental systems. When mediators are correlated, ignoring dependence can reduce statistical efficiency and power for detecting true causal pathways \citep{breiman1997predicting,saccenti2014reflections,lee2017multivariate,lee2020bayesian}.

To address these limitations, \cite{bae2024bayesian} proposed an active mediation pathway identification method based on a Bayesian variable selection framework using a spike-and-slab prior \citep{george1993variable,lee2017multivariate}. Their approach incorporates important dependence structures among high-dimensional mediators by modeling the covariance using a factor-analytic formulation \citep{lee2017multivariate} and applying a Markov random field (MRF) prior. This specification encourages information sharing across correlated mediators and allow pathway selection to be guided by mediator correlation. Additionally, the use of a subsetting prior in the mediator–outcome model, informed by the mediators selected in the exposure–mediator stage, substantially improves computational efficiency.

\begin{figure}[ht]
    \centering
    \includegraphics[width=0.95\linewidth]{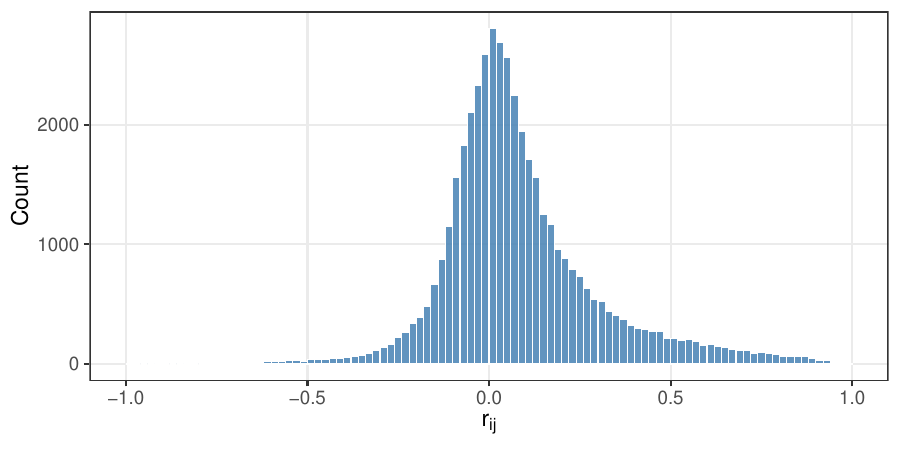}
    \caption{Histogram of pairwise correlations $r_{ij}$ among the 298 candidate metabolites in the HPFS/NHSII application $(1 \le i < j \le 298)$. The mean absolute correlation among metabolite pairs is 0.157, indicating substantial dependence among mediators and motivating correlation-guided pathway selection. 
    Moreover, 13.4\% of metabolite pairs have $|r_{ij}| \ge 0.3$ and 5.7\% have $|r_{ij}| \ge 0.5$, indicating a non-negligible tail of strong correlations.}
    \label{fig:histogram}
\end{figure}

However, this methodology focuses exclusively on continuous outcomes, and both the modeling strategy and computation scheme were developed specifically for that setting. In many studies, outcomes are non-continuous, including binary endpoints such as disease incidence, count outcomes, which are naturally modeled using generalized linear models (GLMs). Extending high-dimensional mediation analysis to GLMs is not a direct plug-in replacement, because the likelihood is no longer Gaussian and the posterior distribution does not admit a closed-form Gibbs sampler. In this paper, we extend their framework by incorporating GLMs for the outcome. The resulting posterior computation is carried out using scalable algorithm based on Hamiltonian Monte Carlo \citep{neal2011mcmc}. We also define target estimands and effect decomposition for non-continuous outcomes, using binary responses as a representative example. Motivated by metabolomics data from substudies within the Health Professionals Follow-up Study (HPFS) and Nurses’ Health Study II (NHSII), we present an application analysis to illustrate the proposed framework in a setting with strongly correlated candidate mediators and a binary outcome. Specifically, we investigate the effect of a dietary pattern score on a binary cardiometabolic risk endpoint, with plasma metabolites considered as potential mediators, where substantial correlations among metabolites motivates modeling approaches that account for mediator dependence (see Figure \ref{fig:histogram}).

The remainder of this paper is organized as follows. Section 2 introduces the high-dimensional mediation model along with the MRF and spike-and-slab priors used for variable selection. Section 3 defines the target estimands and illustrates effect decomposition for non-continuous outcomes, with binary responses presented as a representative example. Section 4 details the proposed computational algorithm, and Section 5 evaluates its empirical performance across diverse simulation scenarios. Section 6 presents an application to metabolomics dataset. We conclude with a summary and discussion in Section 7.

\section{Model Framework and Bayesian Pathway Selection Guided by Mediator Correlation} \label{sec:model}

\begin{figure}[ht]
\centering
\includegraphics[width=0.95\textwidth]{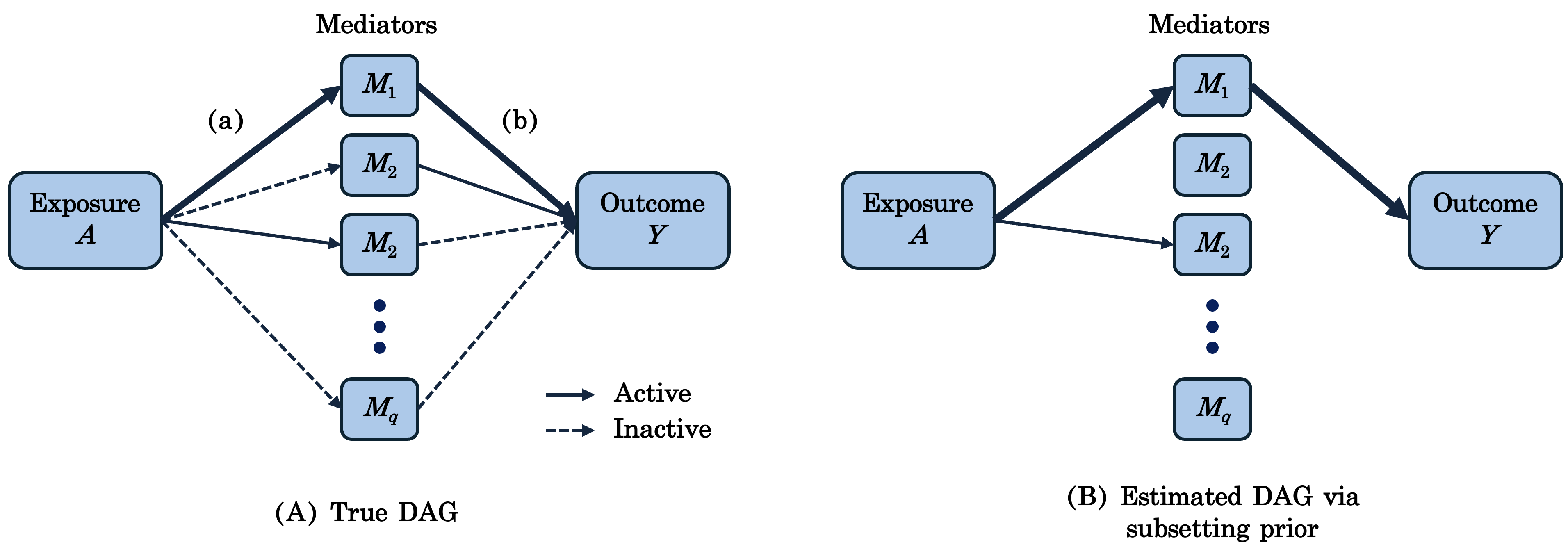}
\caption{True directed acyclic graph (DAG) illustrating $q$-dimensional mediators (A) and the estimated DAG based on the proposed subsetting prior (B). Solid lines represent active pathways (i.e., those with a true causal effect). A causal effect is transmitted through a mediator only when both the (a) exposure–mediator and (b) mediator–outcome pathways are active, as indicated by thick solid lines. Under the subsetting prior, the activity of mediator–outcome pathways is evaluated only for mediators with active exposure–mediator connections, resulting in identical estimated active pathways.
} \label{fig:DAG}
\end{figure}

In the presence of high-dimensional potential mediators, our goal is to identify which mediators transmit the effect of the exposure to the outcome of interest, that is, to determine which mediation pathways are active. As illustrated in Figure \ref{fig:DAG} (A), four possible combinations arise depending on whether the exposure-to-mediator path (a) and the mediator-to-outcome path (b) are active. A mediator is regarded as transmitting an effect only when both paths (a) and (b) are simultaneously active, as indicated by the thick solid line. 

To evaluate the activity of the two pathways, we specify models for both the mediators and the outcome within a Bayesian variable selection framework using spike-and-slab priors. If path (a) is not selected, there is no need to assess path (b), since only pathways with both paths (a) and (b) active can transmit a mediation effect, as shown in Figure \ref{fig:DAG} (A). This motivates a subsetting prior that improves computational efficiency by restricting selection on path (b) to mediators identified as active along path (a) (Figure \ref{fig:DAG} (B)). We assume that mediation pathways are causally independent, without overlapping pathways across mediators, but allow for associational dependence among mediators. To account for such dependencies, we introduce a MRF prior that enables information sharing across correlated mediators and improves pathway selection. In particular, pathway selection is guided by mediator correlation through dependence among inclusion indicators in the prior specification.

Although most existing methods are restricted to continuous outcomes, our framework extends to non-continuous outcomes within the exponential family by modeling the outcome using a GLM. Unlike the continuous outcome setting, the resulting non-Gaussian likelihood alters the posterior updating of the outcome model parameters, which motivates the computational strategy described in Section \ref{sec:posterior}. We demonstrate the proposed method primarily for binary responses, while the modeling framework extends naturally to other outcome types in the exponential family.

\subsection{Notation and model specification} \label{model_eq}

Let $\boldsymbol{X}_i$, $A_i$, $\boldsymbol{M}_i$, and $Y_i$ denote the $p$-dimensional covariate vector, exposure, $q$-dimensional mediator vector, and outcome for subject $i$, respectively. The proposed modeling framework specifies high-dimensional mediator and outcome models as
\begin{align}
\boldsymbol{M}_i &= \boldsymbol{\beta}_0 + \boldsymbol{\tau} A_i + B^\top \boldsymbol{X}_i + \boldsymbol{\epsilon}_M, 
\quad \boldsymbol{\epsilon}_M \sim \mathrm{MVN}(\mathbf{0}, \Sigma), \label{eq:mediator}\\
g(\mu_i) &= \alpha_0 + \boldsymbol{\delta}^\top \boldsymbol{M}_i + \boldsymbol{\alpha}^\top \boldsymbol{X}_i + \alpha_{p+1} A_i. \label{eq:outcome}
\end{align}
We assume $Y_i$ follows an exponential family distribution with mean $\mu_i = E(Y_i | A_i, \boldsymbol{X}_i, \boldsymbol{M}_i)$ and $g(\mu_i)$, a known link function that relates the linear predictor to the conditional mean response. In the mediator model, $\boldsymbol{\beta}_0 = (\beta_{0,1}, \ldots, \beta_{0,q})^\top$ denotes the vector of mediator-specific intercepts, $B \in \mathbb{R}^{p \times q}$ is the matrix of regression coefficients with columns $\boldsymbol{\beta}_j = (\beta_{1,j}, \ldots, \beta_{p,j})^\top$ for $j = 1, \ldots, q$, and $\boldsymbol{\tau} = (\tau_1, \ldots, \tau_q)^\top$ represents the effects of the exposure $A_i$ on the mediators. The covariance matrix $\Sigma$ captures the residual dependence structure among the mediators. In the outcome model, $\boldsymbol{\delta} = (\delta_1, \ldots, \delta_q)^\top$ and $\boldsymbol{\alpha} = (\alpha_1, \ldots, \alpha_p)^\top$ denote the regression coefficients for the mediators and covariates, respectively, and $\alpha_{p+1}$ quantifies the direct effect of the exposure on the outcome.

A key feature of this formulation is the use of the link function $g(\cdot)$, which embeds the outcome model within the GLM framework and thus accommodates any outcome type from the exponential family, 
including continuous, binary, count, or categorical responses. 
This generalization broadens the applicability of the proposed mediation framework and leads to a non-Gaussian likelihood for which posterior computation must be adapted. The resulting computational algorithm is described in Section \ref{sec:posterior}, and implementation details are provided in the Supplementary Materials A.

For illustration, we focus on binary outcomes. Let $\mu_i = \pi_i = \Pr(Y_i = 1)$ and adopt the logit link $g(\mu_i) = \log\{\pi_i / (1 - \pi_i)\}$. The probit link can also be accommodated within the proposed framework and is considered in additional numerical studies. The multivariate mediator model in Eq.~(\ref{eq:mediator}) captures residual correlations among mediators, which are propagated through to the estimation of indirect effects. Under the assumption of causally independent mediation pathways, the outcome model in Eq.~(\ref{eq:outcome}) includes mediators as additive main effects, allowing direct quantification of their individual contributions to the overall mediation effect.

\subsection{Covariance structure and MRF prior}

To incorporate both causally independent pathways and potential correlations among mediators, we adopt a covariance structure that flexibly accommodates dependence while maintaining the assumption of no direct causal relationships between mediators. Specifically, for the covariance matrix $\Sigma$ in Eq.~(\ref{eq:mediator}), we employ a factor analytic (FA) model \citep{lee2017multivariate} of the form
\[
\Sigma = \sigma_\Sigma^2 (\boldsymbol{\lambda} \boldsymbol{\lambda}^\top + I_q),
\]
where $\boldsymbol{\lambda} = (\lambda_1, \ldots, \lambda_q)^\top$ is a $q \times 1$ vector of factor loadings and $I_q$ is the $q \times q$ identity matrix. This FA parameterization captures low-rank covariance structures frequently observed in high-dimensional biological mediators, enabling computational tractability while allowing moderate to strong correlations. This specification employs a single vector-valued latent factor through the loading vector $\boldsymbol{\lambda}$; it can be extended to multiple factors \citep{bartholomew2011latent}, but we focus on a parsimonious formulation that remains computationally tractable in high dimensions. 

This modeling choice preserves the assumption that each mediator lies on an independent causal pathway while permitting non-causal correlations through $\Sigma$. Consequently, the model effectively separates causal from associational structure, ensuring that mediation effects are properly attributed even in the presence of biologically realistic dependence among mediators. While the FA model captures marginal covariance among mediators, additional structural dependence in the context of variable selection is modeled through an MRF prior, which will be described in detail in the subsequent section. This MRF construction allows variable selection to be guided by mediator correlation through dependence among inclusion indicators.

\subsection{Prior specification for variable selection}

To enable the selection of active exposure–mediator and mediator–outcome pathways, we assign spike-and-slab priors to $\tau_j$ and $\delta_j$ for $j = 1, \ldots, q$:
\begin{align*}
    \tau_j \mid \gamma_j, \Sigma &\sim \gamma_j \, \mathrm{Normal}(0, \nu_j^2 \Sigma_{(j,j)}) + (1 - \gamma_j)\, \mathcal{I}_0, \\
    \delta_j \mid \omega_j &\sim \omega_j \, \mathrm{Normal}(0, \psi_j^2) + (1 - \omega_j)\, \mathcal{I}_0,
\end{align*}
where $\gamma_j$ and $\omega_j$ are latent indicators denoting whether the corresponding coefficients belong to the slab or spike component, and $\nu_j^2$ and $\psi_j^2$ are hyperparameters to be specified.

We model dependence among mediators in the selection process via an MRF prior on $\boldsymbol{\gamma}$ \citep{li2010bayesian, lee2017multivariate, bae2024bayesian}. The conditional distributions of the latent indicators are given by
\[
\Pr(\gamma_j = 1 \mid \boldsymbol{\gamma}_{(-j)}, \eta_\gamma)
= \frac{\exp\{\eta_{1\gamma} + \eta_{2\gamma} \sum_{l \in \mathcal{N}_j} r_{jl} \gamma_l\}}
{1 + \exp\{\eta_{1\gamma} + \eta_{2\gamma} \sum_{l \in \mathcal{N}_j} r_{jl} \gamma_l\}},
\]
where $\mathcal{N}_j$ denotes the neighborhood set of mediator $j$, $r_{jl}$ quantifies the strength of association (e.g., correlation) between mediators $j$ and $l$, and $\eta_{1\gamma}, \eta_{2\gamma}$ are MRF hyperparameters controlling sparsity and local smoothness. Here, $\eta_{1\gamma}$ controls the baseline sparsity level, while $\eta_{2\gamma}$ controls the degree to which selection is guided by mediator correlation through the neighborhood structure. When $\eta_{2\gamma}=0$, the prior reduces to independent Bernoulli distribution. For large values of $\eta_{2\gamma}$, the MRF prior can exhibit abrupt changes in the size of the selected set, as noted by \cite{li2010bayesian} and \cite{bae2024bayesian}. In practice, we choose $\eta_{2\gamma}$ within a stable region using a phase-transition diagnostic described in Section \ref{sec:posterior}.

To reflect the multiplicative nature of mediation effects, we further propose a sequential subsetting Bernoulli (SSB) prior, under which activation of $\omega_j$ depends on the inclusion of the corresponding $\gamma_j$:
\[
\Pr(\omega_j = 1 \mid \gamma_j) =
\begin{cases}
    \pi_{\omega}, & \text{if } \gamma_j = 1, \\
    0, & \text{if } \gamma_j = 0,
\end{cases}
\]
where $\pi_{\omega}$ denotes the inclusion probability conditional on an active exposure–mediator link. This hierarchical construction ensures that the mediator–outcome pathway is activated only when the corresponding exposure–mediator effect is present, thereby improving interpretability and computational efficiency.

We assign multivariate normal (MVN) priors to regression coefficients in the mediator and outcome models: $\boldsymbol{\beta}_j^\prime = (\beta_{0,j}, \beta_{1,j}, \ldots, \beta_{p,j})^\top\sim \textrm{MVN}(0, \boldsymbol{\sigma}_{\beta} I_{p+1})$ for $j=1,\ldots q$, $\boldsymbol{\alpha}^\prime=(\alpha_0, \alpha_1, \ldots, \alpha_{p+1})^\top \sim \text{MVN}(\boldsymbol{0}, \boldsymbol{\sigma}_{\alpha} I_{p+2})$ where $\boldsymbol{\sigma}_\beta = diag(\sigma_{\beta_0}, \sigma_{\beta_1}, \ldots, \sigma_{\beta_p})$ and $\boldsymbol{\sigma}_\alpha = diag(\sigma_{\alpha_0}, \sigma_{\alpha_1}, \ldots, \sigma_{\alpha_{p+1}})$. We use standard priors to covariance parameters in the factor analytic specification: $\blambda \sim \textrm{MVN}(\mu_\lambda, h_\lambda \sigma_\Sigma^2 I_q)$ and $\sigma_\Sigma^2 \sim \text{inverse-Gamma}(\nu_0/2, \nu_0 \sigma_0^2 / 2)$ was used. In this case, $\boldsymbol{\sigma}_\beta, \boldsymbol{\sigma}_\alpha, \mu_\lambda, h_\lambda, \nu_0, \sigma_0^2$ are hyperparameters to be specified.

\section{Causal Estimands and Effect Decomposition}
In conventional single-mediator analysis, the primary causal estimands of interest are the \emph{direct effect} and the \emph{indirect effect}, both defined by comparing potential outcomes under varying exposure levels. To extend this framework to a high-dimensional mediation setting, we generalize the potential outcomes framework \citep{rubin1974estimating} as follows. Let $\boldsymbol{M}_i(a) = \left\{M_{1,i}(a), M_{2,i}(a), \cdots, M_{q,i}(a)\right\}^\top$ denote the vector of potential mediator values for subject $i$ when the exposure is set to $A_i = a$. Similarly, let $Y_i(a, \boldsymbol{M}_i(a^\star))$represent the potential outcome under exposure status $A_i = a$ and the mediator vector realized under exposure $A_i = a^\star$. Under this notation, causal estimands in high-dimensional mediation analysis are characterized by contrasts between potential outcomes evaluated under different exposure conditions. Although it is possible to define additional quantities to assess individual mediator-level contributions \citep{vansteelandt2017interventional,bellavia2018decomposition}, such methods require stronger identifying assumptions. We therefore restrict our focus to the contrasts based on the aforementioned quantities. 

For a binary outcome variable $Y_i$, the causal effect of exposure $A_i$ on $Y_i$ can be defined using odds ratios comparing exposure levels $a$ and $a^\star$ \citep{vanderweele2014mediation}. Although target estimands can be expressed via risk differences, we employ the odds ratio formulation as it allows for the direct utilization of the regression structures specified in Eqs (\ref{eq:mediator}) and (\ref{eq:outcome}). The odds ratio of the total effect, representing the change in the odds of $Y_i=1$ when the exposure changes from $a^\star$ to $a$, is defined as:
\begin{eqnarray*}
OR^{TE}_{a,a^\star} =
\frac{P\big(Y_i(a, \boldsymbol{M}_i(a))=1\big)/P\big(Y_i(a, \boldsymbol{M}_i(a))=0\big)}
{P\big(Y_i(a^\star, \boldsymbol{M}_i(a^\star))=1\big)/P\big(Y_i(a^\star, \boldsymbol{M}_i(a^\star))=0\big)}.
\end{eqnarray*}
In mediation analysis, the total effect can be decomposed further into the natural direct effect (DE) and the natural indirect effect (IE). The direct effect, capturing the effect of exposure $A_i$ on $Y_i$ while the mediator vector is held at its counterfactual value under $A_i=a^\star$, and the indirect effect, capturing the change in odds due to the change in mediators from $\boldsymbol{M}_i(a^\star)$ to $\boldsymbol{M}_i(a)$, are defined respectively as:
\begin{eqnarray*}
    OR^{DE}_{a,a^\star} &=&
    \frac{P\big(Y_i(a, \boldsymbol{M}_i(a^\star))=1\big)/P\big(Y_i(a, \boldsymbol{M}_i(a^\star))=0\big)}
    {P\big(Y_i(a^\star, \boldsymbol{M}_i(a^\star))=1\big)/P\big(Y_i(a^\star, \boldsymbol{M}_i(a^\star))=0\big)}, \\
    OR^{IE}_{a,a^\star} &=&
    \frac{P\big(Y_i(a, \boldsymbol{M}_i(a))=1\big)/P\big(Y_i(a, \boldsymbol{M}_i(a))=0\big)}
    {P\big(Y_i(a, \boldsymbol{M}_i(a^\star))=1\big)/P\big(Y_i(a, \boldsymbol{M}_i(a^\star))=0\big)}.
\end{eqnarray*}
These estimands represent the change in the odds of the outcome when the mediating paths are blocked or activated, respectively.
Each of $OR^{DE}_{a,a^\star}$ and $OR^{IE}_{a,a^\star}$ refers to the effect how the odds of outcome $Y_i=1$ change under two exposure levels when blocking the mediating path and through the mediators, respectively. This definition retains the multiplicative decomposition property typical of odds ratios:
$OR^{TE}_{a,a^\star} = OR^{DE}_{a,a^\star} \times OR^{IE}_{a,a^\star}$. Or, it is expressed in the form of addition on the log scale, $\log\big(OR^{TE}_{a,a^\star}\big) = \log\big(OR^{DE}_{a,a^\star}\big) + \log\big(OR^{IE}_{a,a^\star}\big)$. 

Under the model structure assumed in Eqs (\ref{eq:mediator}) and (\ref{eq:outcome}) with $q$-dimensional mediators, the total log-odds effect can be expressed as:
\begin{eqnarray*}
\log\big(OR^{TE}_{a,a^\star}\big) = \alpha + \sum_{j=1}^{q} \tau_j \, \delta_j,
\end{eqnarray*}
where $\alpha$ denotes the log-odds ratio corresponding to the direct effect of the exposure, and $\tau_j \delta_j$ represents the log-odds contribution of the indirect path through the $j$-th mediator. Consequently, $\exp(\tau_j \delta_j)$ represents the odds ratio effect attributed to that specific mediator pathway.

\section{Posterior Computation and Practical Considerations}\label{sec:posterior}

Posterior inference for the proposed model is carried out using a Markov chain Monte Carlo (MCMC) algorithm that builds on the computational strategy of \cite{bae2024bayesian} for continuous outcomes. The key difference is the update of the outcome-model parameters. Under the GLM, the likelihood is non-Gaussian and the corresponding full conditional distributions are not available in closed form. We therefore update the outcome-model coefficients using Metropolis--Hastings steps. To improve mixing after variable selection, we also adopt a refinement step that performs additional joint updates of the nonzero coefficients using Hamiltonian Monte Carlo (HMC) \citep{neal2011mcmc}. This refinement step is used as a computational device to move efficiently along posterior correlation directions among the selected coefficients, and it does not change the target posterior distribution. Full updating steps and implementation details are provided in the Supplementary Materials A. 

As noted in prior work on MRF-based Bayesian variable selection, the smoothness parameter $\eta_{2\gamma}$ can exhibit a sharp phase transition behavior, where small changes in $\eta_{2\gamma}$ lead to abrupt changes in the posterior number of selected mediators \citep{stingo2011incorporating, li2010bayesian, lee2017multivariate, bae2024bayesian}. To select $\eta_{2\gamma}$ within a stable region, we follow the modified posterior simulation approach used in \cite{bae2024bayesian}, which is based on the phase transition diagnostic described by \cite{lee2017multivariate}. Specifically, we run short posterior simulations over a grid of candidate values $\{\eta_g: g=0,\ldots,G,\ \eta_0=0\}$, compute a summary of posterior inclusion behavior of $\gamma$ at each $\eta_g$, and define the phase transition boundary $\eta_{pt}$ as the smallest grid value at which the summary exceeds its value at $\eta_0$ by a prespecified tolerance. We then set $\eta_{2\gamma}$ to the largest grid value below $\eta_{pt}$. 

For pathway selection, we use marginal posterior probabilities of inclusion (PPI) for the joint pathway indicator, ${\rm PPI}_j = P(\gamma_j = 1, \omega_j = 1 \mid \mathcal{D})$. In \cite{bae2024bayesian}, thresholding was performed using an adaptive Bayesian false discovery rate (BFDR) rule \citep{newton2004detecting}. In this paper, we use a fixed cutoff of 0.5, corresponding to the median probability model. This choice was guided by preliminary simulation studies for generalized linear model outcomes, where the 0.5 cutoff yielded more favorable operating characteristics in our settings. We therefore use this rule throughout the numerical studies reported in Sections \ref{sec:simulation} and \ref{section: application}.

To improve computational efficiency, the main sampling engine is implemented in C++. The code is modularized so that it can be extended to other outcome types in the exponential family by replacing only the likelihood component. The full algorithmic code is provided in the published GitHub repository, https://github.com/YounghoBae/BVSmed\_GLM. Computation time was approximately 30 minutes for 10,000 MCMC iterations under a setting with $n = 466, q = 298, p = 3$ on a 3.2 GHz Apple M1 MacBook Pro.

\section{Simulation Studies}\label{sec:simulation}
To explore the finite-sample performance of our proposed Bayesian framework, we conducted numerical studies under various data scenarios. The simulation design is similar to that of the continuous-outcome study in \cite{bae2024bayesian}, with modifications for binary responses and link-function specification.

\subsection{Set-up and data generation}\label{sec:sim_setup}
Five data-generating scenarios were examined. We generated data in Scenarios I--IV with $n = 1,000$  observations, $q = 300$ mediators, and $p = 5$ covariates from the model described in Section \ref{sec:model}. For Scenario V, we used the same dimensions of the empirical application by setting to $(n=466, \;q=298, \;p=3)$. Dependence among mediators was introduced through a single-factor covariance structure. We set the loading parameter $\boldsymbol{\lambda} = 0.35\cdot\mathbf{1}_q$, which yielded an average correlation of roughly 0.1 among mediators (Scenarios I, III, IV). Independence was enforced by setting $\boldsymbol{\lambda} = \mathbf{0}_q$ in Scenario II. Scenario V adopted an empirical covariance structure estimated from the HPFS/NHSII metabolomics data, which reflects a realistic and potentially misspecified dependence pattern. Covariates $\boldsymbol{X}$ were sampled independently from $\text{Normal}(0, 1)$. The exposure variable $A_i$ followed $\text{Normal}(\boldsymbol{l}^\top \boldsymbol{X}_i, 1)$, where $\boldsymbol{l} = (0.5, 0.2, 0.7, 0.4, 0.6)^\top$ for Scenarios I--IV, and $\boldsymbol{l} = (0.5, 0.2, 0.7)^\top$ for Scenario V, inducing moderate confounding between exposure and outcome.

We adopted the same modeling framework introduced in Section \ref{model_eq}, specifying the mediator and outcome models as in (\ref{eq:mediator}) and (\ref{eq:outcome}). For Scenarios I--III and V, the binary outcome was generated under a logistic regression model with $g(\mu_i) = \log\{\mu_i/(1-\mu_i)\}$. Scenario IV examined robustness to link-function misspecification by generating responses from a probit model and fitting the proposed method under (IV-1) a logit link and (IV-2) a probit link. The parameter vectors $\boldsymbol{\tau}$ and $\boldsymbol{\delta}$ controlled the magnitudes of the exposure--mediator and mediator--outcome effects, respectively. We defined $\boldsymbol{\tau} = (-0.12\cdot\mathbf{1}_5, -0.08\cdot\mathbf{1}_5, -0.04\cdot\mathbf{1}_5, 0.04\cdot\mathbf{1}_5, 0.08\cdot\mathbf{1}_5, 0.12\cdot\mathbf{1}_5, \mathbf{0}_{q-30})^\top, \boldsymbol{\delta} = (\boldsymbol{d} \otimes \mathbf{1}_6, \mathbf{0}_{q-30})^\top,$ with $\boldsymbol{d} = (1.5, 2.0, 2.5, 0, 0)^\top$. Scenario III served as the null design by setting all $\tau_j=0$. Intercepts and nuisance parameters were fixed at $\beta_0 = 0.1\cdot\mathbf{1}_q, B = 0.1\cdot\mathbf{1}_{p\times q}, \alpha_0 = -2.5, \boldsymbol{\alpha} = 2\cdot\mathbf{1}_p, \alpha_{p+1} = 2$.

In summary, Scenario I represents the baseline setting with correlated mediators and active pathways; Scenario II enforces independent mediators and serves as a benchmark for methods that use dependence information; Scenario III is the global null configuration; Scenario IV evaluates robustness to link-function misspecification through probit generation; and Scenario V mimics the application setting by using the HPFS/NHSII covariance structure and empirical data dimensions.

\subsection{Specification of hyperparameters and analysis settings}\label{sec:sim_setting}

We compared the proposed framework with two competing models: (1) MVN-MRF-SSB (proposed), which employs a multivariate normal (MVN) distribution for the mediator model and an MRF prior on the latent variable $\boldsymbol{\gamma}$ to account for dependence among mediators in the variable selection stage; (2) MVN-IB-SSB, which retains the MVN mediator model but replaces the MRF prior with independent Bernoulli (IB) priors; and (3) Normal-IB-SSB, which assumes mutually independent mediators. Across all three models, we used the SSB prior for the mediator--outcome inclusion indicators in the outcome model. In the main paper, we focus on comparing MVN-MRF-SSB with MVN-IB-SSB because both models use the same MVN mediator model and differ only in whether mediator dependence is incorporated in the exposure--mediator selection stage, which isolates the contribution of the MRF prior. We report results for Normal-IB-SSB in the Supplementary Material B.

We set $\boldsymbol{\sigma}_\beta = 100\cdot I_{p+1}, \boldsymbol{\sigma}_\alpha = 100\cdot I_{p+2}, h_\lambda=100,\; (\mu_0,\mu_1,\mu_\lambda)=0,\; \nu_0=6,\; \sigma_0^2=1/3.$ The prior inclusion probabilities for $\gamma_j$ and $\omega_j$ were specified to yield prior inclusion probabilities of 0.1. For MVN-MRF-SSB, we selected the MRF smoothness hyperparameter $\eta_{2\gamma}$ using the phase transition diagnostic described in Section \ref{sec:posterior}. 

Each MCMC chain run used 100,000 iterations and discarded the first 50\% as burn-in. Convergence was assessed using trace plots and the Geweke diagnostic \citep{geweke1991evaluating}. Posterior inference for indirect effects was based on marginal PPI for the joint pathway indicator. We evaluated operating characteristics for both the exposure--mediator indicators $\gamma$ and the pathway indicators $\gamma\times\omega$, including true positive rate (TPR), false positive rate (FPR), positive predictive value (PPV), negative predictive value (NPV), and number of variables selected (NVS). Indirect effect estimates were summarized using posterior medians of $\tau_j\delta_j$ across 120 replicates per scenario.

\begin{table}[htp]
    \caption{Operating characteristics$^{\dag}$ (\%) and the number of variables selected (NVS) for exposure-mediator selection ($\gamma$) and active exposure-mediator-outcome pathway selection ($\gamma \times \omega$) in the multivariate normal models with independent Bernoulli and sequential subsetting Bernoulli priors (MVN-IB-SSB) and the proposed Markov random field and sequential subsetting priors (MVN-MRF-SSB) across the five simulation scenarios in Section \ref{sec:sim_setting}.}\label{tab:sim_roc}
    \centering
    \scalebox{0.65}{
    \begin{tabular}{c c c r r c r r c r r c r r}
    \hline
    \multicolumn{2}{c}{} && \multicolumn{5}{c}{MVN-IB-SSB} && \multicolumn{5}{c}{MVN-MRF-SSB} \\
    \cline{4-8}  \cline{10-14}
    \textbf{Scenario} & \textbf{} && \multicolumn{2}{c}{$\gamma$} && \multicolumn{2}{c}{$\gamma \times \omega$} && \multicolumn{2}{c}{$\gamma$} && \multicolumn{2}{c}{$\gamma \times \omega$} \\
    \cline{4-5} \cline{7-8} \cline{10-11} \cline{13-14}
    \multicolumn{2}{c}{} && \textbf{Mean} & \textbf{(SD)} && \textbf{Mean} & \textbf{(SD)} && \textbf{Mean} & \textbf{(SD)} && \textbf{Mean} & \textbf{(SD)} \\
    \hline
                    & TPR &&  59.1 & ( 5.7) &&  59.4 & ( 7.2) &&  71.6 & ( 6.4) &&  70.7 & ( 8.1) \\
                    & FPR &&   0.3 & ( 0.3) &&   1.5 & ( 0.6) &&   2.0 & ( 1.1) &&   0.6 & ( 0.5) \\
                  I & PPV &&  95.7 & ( 4.6) &&  72.8 & ( 9.9) &&  80.8 & ( 8.2) &&  88.9 & ( 8.4) \\
                    & NPV &&  95.6 & ( 0.6) &&  97.4 & ( 0.4) &&  96.9 & ( 0.7) &&  98.2 & ( 0.5) \\
                    & NVS &&  18.5 & ( 1.8) &&  14.8 & ( 1.9) &&  26.9 & ( 3.8) &&  14.4 & ( 1.8) \\ \cline{1-14}
                    & TPR &&  60.0 & ( 5.1) &&  59.4 & ( 7.4) &&  59.9 & ( 5.0) &&  59.1 & ( 7.3) \\
                    & FPR &&   0.3 & ( 0.3) &&   0.5 & ( 0.4) &&   0.3 & ( 0.4) &&   0.5 & ( 0.4) \\
                 II & PPV &&  96.0 & ( 4.0) &&  89.1 & ( 8.1) &&  96.0 & ( 4.8) &&  89.4 & ( 8.4) \\
                    & NPV &&  95.7 & ( 0.5) &&  97.5 & ( 0.4) &&  95.7 & ( 0.5) &&  97.4 & ( 0.4) \\
                    & NVS &&  18.8 & ( 1.6) &&  12.1 & ( 1.8) &&  18.8 & ( 1.7) &&  12.0 & ( 1.9) \\ \cline{1-14}
                    & TPR &&    -- & (  --) &&    -- & (  --) &&    -- & (  --) &&    -- & (  --) \\
                    & FPR &&   0.5 & ( 0.4) &&   0.5 & ( 0.4) &&   0.5 & ( 0.4) &&   0.5 & ( 0.4) \\
                III & PPV &&    -- & (  --) &&    -- & (  --) &&    -- & (  --) &&    -- & (  --) \\
                    & NPV && 100.0 & ( 0.0) && 100.0 & ( 0.0) && 100.0 & ( 0.0) && 100.0 & ( 0.0) \\
                    & NVS &&   1.5 & ( 1.1) &&   1.5 & ( 1.1) &&   1.4 & ( 1.1) &&   1.4 & ( 1.1) \\ \cline{1-14}
                    & TPR &&  59.0 & ( 5.3) &&  59.4 & ( 7.2) &&  71.4 & ( 6.4) &&  70.4 & ( 8.4) \\
                    & FPR &&   0.3 & ( 0.3) &&   1.6 & ( 0.6) &&   2.1 & ( 1.2) &&   0.8 & ( 0.5) \\
               IV-1 & PPV &&  95.8 & ( 4.4) &&  71.1 & ( 8.9) &&  80.3 & ( 8.4) &&  85.9 & ( 9.0) \\
                    & NPV &&  95.6 & ( 0.5) &&  97.4 & ( 0.4) &&  96.9 & ( 0.7) &&  98.1 & ( 0.5) \\
                    & NVS &&  18.5 & ( 1.7) &&  15.1 & ( 1.7) &&  27.0 & ( 4.0) &&  14.9 & ( 2.0) \\ \cline{1-14}
                    & TPR &&  58.9 & ( 5.2) &&  59.4 & ( 6.9) &&  71.6 & ( 5.9) &&  70.7 & ( 7.9) \\
                    & FPR &&   0.3 & ( 0.3) &&   1.1 & ( 0.6) &&   2.0 & ( 1.1) &&   0.6 & ( 0.5) \\
               IV-2 & PPV &&  95.3 & ( 4.5) &&  78.4 & (10.1) &&  80.7 & ( 8.6) &&  88.2 & ( 9.1) \\
                    & NPV &&  95.6 & ( 0.5) &&  97.4 & ( 0.4) &&  96.9 & ( 0.6) &&  98.2 & ( 0.5) \\
                    & NVS &&  18.6 & ( 1.7) &&  13.8 & ( 1.7) &&  27.0 & ( 3.7) &&  14.5 & ( 1.8) \\ \cline{1-14}
                    & TPR &&  29.8 & ( 7.4) &&  29.1 & ( 9.0) &&  81.6 & (19.6) &&  50.2 & (11.2) \\
                    & FPR &&   0.5 & ( 0.8) &&   1.5 & ( 0.5) &&  11.3 & (10.8) &&   1.4 & ( 1.0) \\
                  V & PPV &&  88.2 & (13.4) &&  55.7 & ( 9.9) &&  60.3 & (26.5) &&  72.0 & (14.2) \\
                    & NPV &&  92.7 & ( 0.7) &&  95.6 & ( 0.5) &&  98.0 & ( 2.1) &&  96.9 & ( 0.7) \\
                    & NVS &&  10.4 & ( 3.1) &&   9.4 & ( 2.5) &&  54.6 & (34.1) &&  13.1 & ( 3.9) \\ \hline
                    \multicolumn{14}{l}{\footnotesize NOTE: Throughout values are based on results from 120 simulated datasets.}\\    
                    \multicolumn{14}{l}{\footnotesize $\dag$ true positive rate (TPR), false positive rate (FPR)} \\     
                    \multicolumn{14}{l}{\footnotesize ~ negative predictive value (NPV), and positive predictive value (PPV)} \\     
                    \multicolumn{14}{l}{\footnotesize$^{\ddag}$ TPR and PPV are not presented, as there are no active pathways.}
    \end{tabular}}
\end{table}

\begin{table}[htp] 
    \caption{Posterior summaries of mediator-specific indirect effects (IEs)$^\dag$ and marginal posterior probabilities of inclusion (PPI) under Scenario I, comparing the multivariate normal models with independent Bernoulli and sequential subsetting Bernoulli priors (MVN-IB-SSB) and the proposed Markov random field and sequential subsetting priors (MVN-MRF-SSB).} \label{tab:sim_est}
    \centering
    \scalebox{0.64}{
    \begin{tabular}{r r r r c r r c r r c r r r r c r r c r r}
    \hline
    &\multicolumn{3}{c}{} && \multicolumn{2}{c}{MVN-IB-SSB} && \multicolumn{2}{c}{MVN-MRF-SSB} && &  \multicolumn{3}{c}{} && \multicolumn{2}{c}{MVN-IB-SSB} && \multicolumn{2}{c}{MVN-MRF-SSB} \\
     \cline{6-7}\cline{9-10}\cline{17-18}\cline{20-21} 
    &\multicolumn{3}{c}{True} && \multicolumn{2}{c}{IE$_j$${^\ddag}$} && \multicolumn{2}{c}{IE$_j$} && &  \multicolumn{3}{c}{True} && \multicolumn{2}{c}{IE$_j$} && \multicolumn{2}{c}{IE$_j$} \\
    $j$ & \textbf{$\tau_j$} & \textbf{$\delta_j$} & IE$_j$ && PM (SD) & $\overline{\text{PPI}}$ && PM (SD) & $\overline{\text{PPI}}$ && $j$ & $\tau_j$ & $\delta_j$ & IE$_j$ && PM (SD) & $\overline{\text{PPI}}$ && PM (SD) & $\overline{\text{PPI}}$ \\ \cline{1-10} \cline{12-21}
    1  & -0.12 & 1.5 & -0.36 && -0.21 (0.07) & 0.98 && -0.24 (0.09) & 0.97 && 16 &  0.04 & 1.5 &  0.12 &&  0.10 (0.05) & 0.17 && 0.10 (0.06) & 0.35 \\
    2  & -0.12 & 2.0 & -0.48 && -0.28 (0.09) & 0.98 && -0.33 (0.12) & 0.99 && 17 &  0.04 & 2.0 &  0.16 &&  0.13 (0.05) & 0.12 && 0.12 (0.07) & 0.31 \\
    3  & -0.12 & 2.5 & -0.60 && -0.31 (0.09) & 0.99 && -0.38 (0.13) & 1.00 && 18 &  0.04 & 2.5 &  0.20 &&  0.15 (0.08) & 0.10 && 0.14 (0.10) & 0.26 \\
    4  & -0.12 & 0.0 &  0.00 && -0.05 (0.05) & 0.55 && -0.04 (0.06) & 0.28 && 19 &  0.04 & 0.0 &  0.00 &&  0.02 (0.03) & 0.07 && 0.01 (0.03) & 0.08 \\
    5  & -0.12 & 0.0 &  0.00 && -0.06 (0.05) & 0.58 && -0.04 (0.06) & 0.28 && 20 &  0.04 & 0.0 &  0.00 &&  0.03 (0.03) & 0.11 && 0.02 (0.03) & 0.11 \\
    6  & -0.08 & 1.5 & -0.24 && -0.15 (0.06) & 0.70 && -0.15 (0.07) & 0.79 && 21 &  0.08 & 1.5 &  0.24 &&  0.14 (0.05) & 0.65 && 0.16 (0.06) & 0.82 \\
    7  & -0.08 & 2.0 & -0.32 && -0.19 (0.06) & 0.70 && -0.22 (0.08) & 0.85 && 22 &  0.08 & 2.0 &  0.32 &&  0.20 (0.07) & 0.73 && 0.22 (0.08) & 0.86 \\
    8  & -0.08 & 2.5 & -0.40 && -0.23 (0.08) & 0.72 && -0.27 (0.11) & 0.87 && 23 &  0.08 & 2.5 &  0.40 &&  0.23 (0.08) & 0.69 && 0.26 (0.11) & 0.85 \\
    9  & -0.08 & 0.0 &  0.00 && -0.04 (0.04) & 0.38 && -0.03 (0.05) & 0.26 && 24 &  0.08 & 0.0 &  0.00 &&  0.04 (0.04) & 0.36 && 0.02 (0.05) & 0.27 \\
    10 & -0.08 & 0.0 &  0.00 && -0.05 (0.04) & 0.42 && -0.04 (0.04) & 0.28 && 25 &  0.08 & 0.0 &  0.00 &&  0.04 (0.04) & 0.42 && 0.03 (0.05) & 0.28 \\
    11 & -0.04 & 1.5 & -0.12 && -0.09 (0.06) & 0.12 && -0.10 (0.06) & 0.29 && 26 &  0.12 & 1.5 &  0.36 &&  0.22 (0.07) & 0.98 && 0.25 (0.09) & 0.96 \\
    12 & -0.04 & 2.0 & -0.16 && -0.12 (0.06) & 0.11 && -0.12 (0.08) & 0.29 && 27 &  0.12 & 2.0 &  0.48 &&  0.27 (0.09) & 0.98 && 0.32 (0.11) & 1.00 \\
    13 & -0.04 & 2.5 & -0.20 && -0.15 (0.08) & 0.08 && -0.15 (0.10) & 0.28 && 28 &  0.12 & 2.5 &  0.60 &&  0.32 (0.09) & 0.97 && 0.38 (0.12) & 0.99 \\
    14 & -0.04 & 0.0 &  0.00 && -0.03 (0.03) & 0.07 && -0.01 (0.04) & 0.10 && 29 &  0.12 & 0.0 &  0.00 &&  0.05 (0.06) & 0.57 && 0.04 (0.07) & 0.28 \\
    15 & -0.04 & 0.0 &  0.00 && -0.03 (0.03) & 0.07 && -0.01 (0.03) & 0.10 && 30 &  0.12 & 0.0 &  0.00 &&  0.06 (0.05) & 0.57 && 0.04 (0.06) & 0.29 \\
    \hline
    &  \multicolumn{3}{c}{True} && DE${*}$ & && DE & \multicolumn{11}{c}{\textbf{}}\\
    &  \multicolumn{3}{c}{DE}      && \multicolumn{2}{c}{PM (SD)}      && \multicolumn{2}{c}{PM (SD)}      & \multicolumn{11}{c}{\textbf{}}\\ \cline{1-10}
    &  \multicolumn{3}{c}{4.00}    && \multicolumn{2}{c}{1.85 (0.44)} && \multicolumn{2}{c}{2.36 (0.65)}  & \multicolumn{11}{c}{} \\ \hline
    \multicolumn{21}{l}{\footnotesize NOTE: Throughout values are based on results from 120 simulated datasets.} \\
    \multicolumn{21}{l}{\footnotesize ~~~~~~~~~~ Results are shown for the first 30 mediators ($j = 1, \ldots, 30$) that are causally affected by the exposure.} \\
    \multicolumn{21}{l}{\footnotesize $\dag$ IE: indirect effect, defined as $2 \tau \delta$, DE: direct effect, defined as $2\alpha_{p+1}$.} \\
    \multicolumn{21}{l}{\footnotesize ~ The value ``2" is the exposure contrast between the 75th and 25th percentiles of the exposure distribution ($1 - (-1) = 2$).} \\
    \multicolumn{21}{l}{\footnotesize $\ddag$ The medians of the posterior means (PM) and posterior standard deviation (SD) of IE$_{j}$ conditioning on $\gamma_{j}=1$ and $\omega_{j}=1$.} \\
    \multicolumn{21}{l}{\footnotesize ~ The medians of the posterior means of PPI ($\gamma_j\times\omega_j$) are computed.} \\
    \multicolumn{21}{l}{\footnotesize$**$ The medians of the posterior means (PM) and posterior standard deviation (SD) of DE are computed.} 
    \end{tabular}}
\end{table}

\subsection{Results}
Table \ref{tab:sim_roc} summarizes variable selection performance across scenarios. For Scenario I, which represents correlated mediators with active pathways, the proposed MVN-MRF-SSB showed clear gains relative to MVN-IB-SSB. The mean TPR for the exposure--mediator indicators $\gamma$ increased from 59\% to 72\%, and the pathway-level TPR for $\gamma\times\omega$ increased from 59\% to 71\%. At the pathway level, MVN-MRF-SSB also improved PPV from 73\% to 89\%, while maintaining a low FPR (0.6\%). These results indicate that the MRF prior can use mediator correlation information to improve detection of exposure-mediator links, which then translates into improved recovery of active pathways under the sequential subsetting rule.

For Scenario II and III provide benchmark settings. When mediators were independent (Scenario II), both MVN-MRF-SSB and MVN-IB-SSB showed nearly identical performance for both $\gamma$ and $\gamma\times\omega$. Under the global null (Scenario III), both approaches controlled false discoveries, with FPR around 0.5\% and NPV at 100\%. These results suggest that incorporating dependence through the MRF prior does not introduce spurious pathway detection in settings where dependence information is absent or where signals are not present.

Scenario IV aimed to evaluate robustness to link-function misspecification by generating binary outcomes under a probit model and fitting the methods under two links: a logit link (Scenario IV-1) and a probit link (Scenario IV-2). For pathway selection ($\gamma\times\omega$), MVN-MRF-SSB outperformed MVN-IB-SSB under both fitted links, and the operating characteristics were similar across Scenario IV-1 and Scenario IV-2. In particular, the pathway-level TPR for MVN-MRF-SSB was 70\% in Scenario IV-1 and 71\% in Scenario IV-2, compared with 59\% under MVN-IB-SSB in both settings. The pathway-level PPV for MVN-MRF-SSB was 86\% in Scenario IV-1 and 88\% in Scenario IV-2, compared with 71\% and 78\% under MVN-IB-SSB, respectively. These results indicate that the improvement from incorporating mediator dependence through the MRF prior is maintained under moderate link-function misspecification.

Scenario V mimics the real data application by using the empirical covariance structure and data dimensions from the HPFS/NHSII metabolomics study. This setting was the most challenging because it combines strong and heterogeneous mediator dependence with a smaller sample size. 
In this scenario, MVN-MRF-SSB improved pathway-level sensitivity relative to MVN-IB-SSB (TPR 50.2\% versus 29.1\%) and increased PPV (72\% versus 56\%). Although the MRF prior increased the false positive rate for $\gamma$ (11\%), the sequential subsetting rule reduced false positives at the pathway level, yielding a pathway-level FPR of 1.4\%. This pattern suggests that the MRF component improves sensitivity in the exposure--mediator stage, while the sequential subsetting rule helps maintain sparsity for pathway selection.

Table \ref{tab:sim_est} summarizes mediator-specific indirect effect estimation and further illustrates the mechanism behind the gains in Scenario I. For strong signals (e.g., $|\tau_j|=0.12$ with $\delta_j\ge 1.5$), both methods yielded PPI values near 1. For moderate signals ($|\tau_j|=0.08$ with $\delta_j\in\{1.5,2.0,2.5\}$), MVN-MRF-SSB yielded higher PPI values than MVN-IB-SSB, for example for mediators 6--8 and 21--23. For inactive mediators, MVN-MRF-SSB generally yielded smaller PPI values than MVN-IB-SSB in this scenario, which is consistent with the improved pathway-level PPV in Table~\ref{tab:sim_roc}. Additional mediator-specific summaries for Scenarios II--V are provided in the Supplementary Material B.

Overall, the simulation results indicate that MVN-MRF-SSB yields improved pathway recovery when mediators are correlated, while maintaining comparable performance in the independence benchmark and strict error control under the global null. In addition, the method shows stable behavior under link-function misspecification and retains improved sensitivity in an application-mimicking setting with an empirical covariance structure.

\section{Application to Metabolomics Data} \label{section: application}
\subsection{Data description}
We illustrate the proposed method using metabolomics data from substudies within the HPFS and NHSII, with a sample of $n=466$ participants. Our goal is to investigate whether the association between adherence to a Mediterranean-style dietary pattern and insulin resistance is mediated through plasma metabolites. The outcome in the main analysis is a triacylglycerol-to-HDL-cholesterol ratio (TG/HDL-C), a commonly used marker of insulin resistance. A cutoff of $\geq3.5$ is used to dichotomize insulin resistance---individuals at or above this value are considered insulin resistant \citep{mclaughlin2005there}. The exposure is the alternate Mediterranean diet score, a dietary pattern score with higher values indicating greater adherence to a Mediterranean-style diet. We consider $q=298$ plasma metabolites as candidate mediators and adjust for $p=3$ baseline covariates. Metabolite measurements were preprocessed using transformation and standardization steps, and missing values were imputed using a random forest approach. We also conducted an additional analysis using a binary composite cardiometabolic disease risk outcome (CMD). Further data description and the full results for the CMD analysis are provided in the Supplementary Materials C.

\subsection{Hyperparameter specification and analysis settings}
We fitted the proposed MVN-MRF-SSB model and two competing models (MVN-IB-SSB and Normal-IB-SSB) to the HPFS/NHSII data. We adjusted for age, sex, and body mass index as baseline covariates ($p=3$) and did not perform variable selection on these covariates. Hyperparameters were set to the same values used in the simulation studies. Posterior computation was based on three independent MCMC chains, each run for a total of 300,000 iterations, with the first half discarded as burn-in. Convergence was assessed using trace plots and Geweke diagnostics for representative parameters.

\begin{table}[htp]
    \centering
    \caption{Indirect effect estimates (IEs)$^\dag$ for the effect of adherence to the Mediterranean diet on insulin resistance through selected metabolite pathways under MVN-IB-SSB and MVN-MRF-SSB, with estimated diet--metabolite effects ($\tau$).}\label{tab:data_est1}
    \scalebox{0.7}{
    \begin{tabular}{c c c c c c c c c c c c c}
    \hline
    && \multicolumn{5}{c}{MVN-IB-SSB} && \multicolumn{5}{c}{MVN-MRF-SSB} \\
    \cline{3-7} \cline{9-13}
    && \multicolumn{2}{c}{$\tau$} && \multicolumn{2}{c}{IE} && \multicolumn{2}{c}{$\tau$} && \multicolumn{2}{c}{IE} \\
    \textbf{Metabolites}$^\ddag$ && \textbf{PM (SD)}$^*$ & $\overline{\textbf{PPI}}$ && \textbf{PM (SD)}$^{**}$ & $\overline{\textbf{PPI}}$ && \textbf{PM (SD)}$^*$ & $\overline{\textbf{PPI}}$ && \textbf{PM (SD)}$^{**}$ & $\overline{\textbf{PPI}}$ \\
    \hline
            TG(60:12) &&  0.21 (0.03) & 1.00 &&  0.07 (0.26) & 0.55 &&            - &    - &&            - &    - \\
             TG(49:3) && -0.29 (0.03) & 1.00 &&  0.29 (0.51) & 0.54 &&            - &    - &&            - &    - \\
PC(P-38:3)/PC(O-38:4) && -0.27 (0.03) & 1.00 &&  0.49 (0.22) & 0.85 && -0.27 (0.03) & 1.00 &&  0.73 (0.33) & 0.56 \\
             PC(40:9) &&  0.15 (0.03) & 0.96 &&  0.32 (0.23) & 0.73 &&            - &    - &&            - &    - \\
             PC(38:6) &&  0.15 (0.03) & 0.95 && -0.08 (0.27) & 0.54 &&            - &    - &&            - &    - \\
             DG(38:4) &&            - &    - &&            - &    - && -0.12 (0.03) & 1.00 &&  0.30 (0.14) & 0.78 \\
             DG(36:2) &&            - &    - &&            - &    - && -0.10 (0.03) & 1.00 && -0.40 (0.21) & 0.60 \\
      Cer(d18:1/22:0) &&            - &    - &&            - &    - && -0.14 (0.03) & 1.00 && -0.41 (0.22) & 0.70 \\
             CE(22:6) &&  0.22 (0.03) & 0.90 && -0.49 (0.21) & 0.80 &&            - &    - &&            - &    - \\
             CE(20:5) &&  0.20 (0.03) & 0.99 && -0.33 (0.20) & 0.71 &&            - &    - &&            - &    - \\
             CE(18:0) && -0.16 (0.03) & 0.60 && -0.23 (0.09) & 0.53 &&            - &    - &&            - &    - \\
    \hline
    && \multicolumn{2}{c}{} && \multicolumn{2}{c}{DE$^{***}$} && \multicolumn{2}{c}{} && \multicolumn{2}{c}{DE} \\
    \multicolumn{5}{c}{} & \multicolumn{2}{c}{PM (SD)} && \multicolumn{3}{c}{} & \multicolumn{2}{c}{PM (SD)} \\
    \cline{6-7} \cline{12-13}
    \multicolumn{5}{c}{} & \multicolumn{2}{c}{-0.74 (0.36)} && \multicolumn{3}{c}{} & \multicolumn{2}{c}{-0.85 (0.63)} \\
    \hline
    \multicolumn{13}{l}{\footnotesize NOTE: We adjusted for participants' age, sex, and body mass index, potential confounders, but did not perform variable selection on them.}\\
    \multicolumn{13}{l}{\footnotesize$\dag$ IE: indirect effect, defined as $1.03 \tau \delta$.}\\
    \multicolumn{13}{l}{\footnotesize~~~~~The value ``1.03" is the exposure contrast between the 70th and 30th percentiles of the exposure distribution ($0.49-(-0.54)=1.03$).}\\
    \multicolumn{13}{l}{\footnotesize $\ddag$ TG: triacylglycerol, PC: Phosphatidylcholine, DG: Diacylglycerol, Cer: Ceramide, CE: cholesterol ester}\\
    \multicolumn{13}{l}{\footnotesize $*$ Posterior mean (PM) and posterior standard deviation (SD) of $\tau$ conditioning on $\gamma=1$}\\
    \multicolumn{13}{l}{\footnotesize $**$ Posterior mean (PM) and posterior standard deviation (SD) of $1.03\tau\delta$ conditioning on $\gamma=\omega=1$} \\
    \multicolumn{13}{l}{\footnotesize ${***}$ Posterior mean (PM) and posterior standard deviation (SD) of DE, defined as $1.03\alpha_{p+1}$}
    \end{tabular}}
\end{table}

\subsection{Results}

We present results for the insulin resistance outcome (TG/HDL-C) using MVN-IB-SSB and MVN-MRF-SSB in the main paper. Results for Normal-IB-SSB and the CMD outcome analysis are provided in the Supplementary Materials C.

Table \ref{tab:data_est1} summarizes indirect effect estimates and posterior inclusion probabilities for pathways selected by MVN-IB-SSB and MVN-MRF-SSB. Overall, MVN-MRF-SSB selected fewer mediators than MVN-IB-SSB, while still identifying mediators with plausible links to cardiometabolic pathways. Notably, the overlap between the two multivariate mediator models consisted of a single phosphatidylcholine feature (PC(P-38:3)/PC(O-38:4)), whereas the remaining selected mediators differed across methods. MVN-IB-SSB selected several triglyceride and cholesterol ester species, while MVN-MRF-SSB selected two diacylglycerol features annotated as DG(38:4) and DG(36:2) and one ceramide species (Cer(d18:1/22:0)) in addition to PC(P-38:3)/PC(O-38:4).

Among the MVN-MRF-SSB findings, Cer(d18:1/22:0) was selected only by the proposed model. Ceramides have been studied as lipid biomarkers associated with cardiometabolic risk, and the negative indirect effect through Cer(d18:1/22:0) is directionally consistent with a pathway in which higher adherence to a Mediterranean-style diet is associated with lower ceramide levels and lower insulin resistance \citep{laaksonen2016plasma}. In contrast, the two DG species exhibited indirect effects in opposite directions. Given this heterogeneity across DG species, we interpret the DG results as identification of candidate DG-related pathways rather than evidence for a single, class-level mechanistic direction. For PC(P-38:3)/PC(O-38:4), the estimated indirect effect was positive despite a negative exposure-to-metabolite pathway, illustrating that individual mediated components may operate in different directions within the overall effect decomposition.

\section{Discussion}

In this paper, we propose correlation-guided Bayesian pathway selection for high-dimensional mediation analysis under generalized linear models. Extending the framework of \citet{bae2024bayesian} from continuous outcomes to GLM outcomes is not a plug-in change, since the likelihood is non-Gaussian and posterior computation requires different updating strategies (see Section \ref{sec:posterior} and Supplementary Materials A). Our approach performs correlation-guided selection through the MRF prior on inclusion indicators and the SSB prior for mediator--outcome selection, while adapting posterior computation for non-continuous outcomes. We illustrate the proposed framework primarily for binary responses, although the modeling strategy extends to other outcome types in the exponential family. The accompanying software \texttt{BVSmed} is implemented in a modular manner so that other GLM outcome types can be handled by replacing only the likelihood component.

Our simulation studies suggest that incorporating mediator dependence through the MRF prior can improve operating characteristics in correlated-mediator settings, yielding higher power while maintaining low false positive rates. In the HPFS/NHSII application, MVN-MRF-SSB selected a smaller set of mediators than MVN-IB-SSB, which is consistent with the intended role of correlation-guided selection in reducing spurious discoveries when candidate mediators are strongly correlated. At the same time, because the true mediating set is unknown in observational data, the selected mediators should be interpreted as candidates for follow-up rather than confirmed causal mechanisms.

Several practical considerations and limitations merit discussion. First, additional improvements in computational efficiency, especially for the mediator-model component, may be needed for routine use in large-scale applications. In the current implementation, the mediation pathway parameters (e.g., $\tau$ and $\delta$) are updated once per iteration, and this may not always yield sufficiently rapid mixing when posterior dependence among selected coefficients is strong. In the application analysis, longer runs (e.g., 300,000 iterations) appeared adequate based on trace plots and convergence diagnostics, whereas in some simulation settings with fewer iterations the diagnostics were less stable. This suggests that further computational strategies could be beneficial, such as increasing the frequency of refinement updates for selected coefficients and adopting more aggressive block updates that better reflect posterior correlation structure.

Second, the current calibration of the MRF smoothness parameter (i.e., the grid-based procedure for $\eta_{2\gamma}$) may depend on subjective choices, such as the diagnostic summary used to identify a stable region and the tolerance used to define the phase transition boundary. This issue can become more pronounced as the number of mediators and the number of truly active mediators vary across studies. Developing a more automated calibration rule, for example a data-adaptive criterion that targets stable posterior inclusion behavior or predictive performance, may improve robustness and reproducibility \citep{clark2025nutritionally}. In addition, pathway selection also requires a decision rule for translating posterior inclusion probabilities into selected mediators. While \cite{bae2024bayesian} adopted an adaptive BFDR rule, our preliminary simulation studies under GLM outcomes suggested that the fixed cutoff of 0.5 yielded more favorable operating characteristics in our settings, and we therefore used it throughout the numerical studies. It remains of interest to further characterize when BFDR-type decision rules offer advantages for GLM outcomes, particularly under alternative signal strengths and correlation patterns.

Finally, the proposed SSB prior improves interpretability and computational efficiency by restricting mediator--outcome selection to mediators supported in the exposure--mediator stage. This design reflects the definition of an active mediation pathway, which requires both the exposure--mediator and mediator--outcome components to be present. At the same time, in settings where signals are weak or the exposure--mediator stage is conservative, it could be of interest to consider extensions that relax the strict subsetting rule, for example by allowing a small probability of mediator--outcome activation even when the exposure--mediator indicator is inactive, or by introducing joint updating strategies that more directly propagate information across stages. 

Overall, the proposed framework provides a practical and scalable approach to correlation-guided Bayesian pathway selection for high-dimensional mediation analysis with non-continuous outcomes. By combining dependence-aware priors with posterior computation tailored to GLMs, the method enables coherent uncertainty quantification and mediator selection in settings where candidate mediators are strongly correlated. We expect the proposed approach to be useful in a broad range of modern applications with high-dimensional mediators and non-Gaussian outcomes.

\section*{Funding}

This work was supported by the NIH of USA (R01GM126257, U01CA167552, U01CA176726) and the NRF grant of South Korea (RS-2025-00554477, RS-2024-00407300).

\section*{Supporting Information}
Supplementary material is available online.

\bibliographystyle{apalike}
\bibliography{BVSmediation}

\end{document}